\documentclass{nature}

\usepackage{amssymb,amsmath,graphicx,color,ulem,soul,multirow}
\usepackage[font=footnotesize]{caption}

\newcommand{\beq}{\begin{equation}}
\newcommand{\eeq}{\end{equation}}
\newcommand{\bea}{\begin{eqnarray}}
\newcommand{\eea}{\end{eqnarray}}
\newcommand{\nn}{{\nonumber}}

\title{Stringent axion constraints with Event Horizon 
Telescope polarimetric measurements of M87$^\star$}

\begin{document}

\author{Yifan Chen$^{1}$, 
Yuxin Liu$^{2}$,
Ru-Sen Lu$^{3,4,5}$,
Yosuke Mizuno$^{6,7}$,
Jing Shu$^{1,8,9,10,11,12}$,
Xiao Xue$^{1,8}$,
Qiang Yuan$^{13,14,10}$,
Yue Zhao$^{15}$}

\maketitle

\begin{affiliations}
\small
\item{CAS Key Laboratory of Theoretical Physics, Institute of Theoretical
Physics, Chinese Academy of Sciences, Beijing 100190, China}
\item{Department of Physics, Tsinghua University, Beijing 100084, China}
\item{Shanghai Astronomical Observatory, Chinese Academy of Sciences, Shanghai 200030, China}
\item{Key Laboratory of Radio Astronomy, Chinese Academy of Sciences, Nanjing 210008, China}
\item{Max-Planck-Institut f\"ur Radioastronomie, Auf dem H\"ugel 69, D-53121 Bonn, Germany}
\item{Tsung-Dao Lee Institute and School of Physics and Astronomy, Shanghai Jiao Tong University, Shanghai, 200240, China}
\item{Institute for Theoretical Physics, Goethe University Frankfurt, Frankfurt am Main, 60438, Germany}
\item{School of Physical Sciences, University of Chinese Academy of Sciences, Beijing 100049, China}
\item{CAS Center for Excellence in Particle Physics, Beijing 100049, China}
\item{Center for High Energy Physics, Peking University, Beijing 100871, China}
\item{School of Fundamental Physics and Mathematical Sciences, Hangzhou Institute for Advanced Study,
University of Chinese Academy of Sciences, Hangzhou 310024, China}
\item{International Center for Theoretical Physics Asia-Pacific, Beijing/Hangzhou, China}
\item{Key Laboratory of Dark Matter and Space Astronomy, Purple Mountain
Observatory, Chinese Academy of Sciences, Nanjing 210023, China}
\item{School of Astronomy and Space Science, University of Science and
Technology of China, Hefei 230026, China}
\item{Department of Physics and Astronomy, University of Utah, Salt Lake City, UT 84112, USA}
\end{affiliations}

\begin{abstract}


The hitherto unprecedented angular resolution of the Event Horizon Telescope (EHT) has created exciting opportunities in the search for new physics. 
Recently, the linear polarization of radiation emitted near the supermassive black hole M87$^\star$ was measured on four separate days, precisely enabling tests of the existence of a dense axion cloud produced by a spinning black hole. 
The presence of an axion cloud leads to a frequency-independent oscillation in the electric vector position angle (EVPA) of this linear polarization. For a nearly face-on M87$^\star$, this oscillation in the EVPA appears as a propagating wave along the photon ring. 
In this paper, we leverage the azimuthal distribution of EVPA measured by the EHT to study the axion-photon coupling. 
We propose a novel differential analysis procedure to reduce the astrophysical background, and derive stringent constraints on the existence of axions in the previously unexplored mass window $\sim (10^{-21}-10^{-20})$~eV.

\end{abstract}

\section{Introduction}

Via the technique of very long baseline interferometry, the Event Horizon Telescope (EHT) reaches an unprecedented spatial resolution in the radio-frequency band, imaging the horizon-scale structure of the supermassive black hole (SMBH) M87$^\star$\cite{Akiyama:2019cqa,Akiyama:2019bqs,
Akiyama:2019eap,Akiyama:2019fyp}. 
Such an achievement provides unique opportunities to study physics 
in extreme conditions. In particular, the EHT measures electromagnetic 
radiation at a wavelength of 1.3 millimeter, which is optimal for studying 
the magnetic field around such black holes via polarization measurements. 
Such measurements have been performed very recently\cite{EHTP}, revealing the 
magnetic field structure near the event horizon\cite{EHTM}.

Beyond the astrophysics in the vicinity of an SMBH, such as the accretion 
flow and jet, polarization measurements can also help us probe 
new physics beyond the Standard Model of particle physics. 
The axion is one of the best-motivated candidates for physics beyond the Standard Model, arising initially as a solution to the strong CP  (charge-conjugation and parity)  problem\cite{Peccei:1977hh,Peccei:1977ur,Weinberg:1977ma,Wilczek:1977pj}. 
Axion-like particles are furthermore generically predicted in fundamental theories 
with extra dimensions\cite{Arvanitaki:2009fg}, and are excellent 
dark matter candidates\cite{Preskill:1982cy,Abbott:1982af,Dine:1982ah}.
In Ref.\,\cite{Chen:2019fsq}, the authors proposed that the temporal and 
spatial variations of the electric vector position angle (EVPA) around 
the photon ring of an SMBH can be used to study the existence of an axion cloud 
produced by the superradiance 
mechanism\cite{Penrose:1971uk,ZS,Press:1972zz}, thanks to the axion-induced birefringence effect\cite{Carroll:1989vb,Harari:1992ea}.
The search is sensitive to axion clouds with non-negligible self-interaction\cite{Yoshino:2012kn,Yoshino:2013ofa,Yoshino:2015nsa,Baryakhtar:2020gao}, which is complimentary to constraints from black hole spin measurements\cite{Arvanitaki:2010sy}.

More explicitly, the EHT collaboration presented fiducial polarimetric images for four days (April 5/6/10/11 2017) in Ref.\,\cite{EHTP}, along with the azimuthal distribution of the EVPA, including systematic uncertainties, for two days (April 5/11, 2017). Based on the search strategy 
proposed in Ref.\,\cite{Chen:2019fsq} and the information provided in 
Ref.\,\cite{EHTP}, we adopt an improved search strategy using differential EVPA in the time domain. Such a method can effectively reduce the uncertainties from the astrophysical background. In addition, we include more realistic modelling of the accretion flow and radiative transfer calculation of polarized light. 

We demonstrate that the newly released EHT polarimetric measurements 
provide a powerful probe for studying the axion-photon coupling for 
axions masses near $10^{-20}$~eV. Interesting constraints, surpassing those from other measurements, can already be produced with the published EHT results. The sensitivity can be also significantly improved when more detailed information becomes available.

\section{Axion Cloud-Induced Birefringence Effect}

If the Compton wavelength of an ultralight boson, e.g., the axion in this study, is comparable to the gravitational radius of a rapidly spinning Kerr black hole, a bound state between the axions and the black hole can be spontaneously generated by extracting the black hole's rotational energy\cite{Penrose:1971uk,ZS,Press:1972zz}.
The growth of such a bound state is exponential. This is the so-called superradiance mechanism.
Such a bound state is a close analog to a hydrogen atom, with discrete energy levels and a `fine structure constant' $\alpha \equiv r_g/\lambda_c$, where $r_g$ is the gravitational radius of the black hole and $\lambda_c$ is the reduced Compton wavelength of the axion. The nontrivial self-interaction of the axion can halt the exponential growth, and the peak value of the axion field may become comparable to the decay constant $f_a$\cite{Yoshino:2012kn,Yoshino:2013ofa,Yoshino:2015nsa,Baryakhtar:2020gao}. In such a scenario, the axion cloud around the black hole may lead to a region with the highest possible axion density anywhere in the universe.

{Analogously to the hydrogen atom, the axion cloud can populate states with differing angular momentum quantum numbers, $(l,m)$. 
The mode with $(l,m) = (1,1)$ is the lowest energy state amongst those satisfying the superradiant condition. Out of all possible states populated through superradiance, it has a peak located closest to the
horizon of the BH. For an extreme Kerr black hole, $\alpha$ needs to be smaller than $0.5$ in order for the $(1,1)$ state to be populated.  In this study we focus on the SMBH M87$^\star$, and consider the axion mass window from  $2 \times 10^{-21}$ eV to $10^{-20}$ eV, which gives $\alpha$ from $0.1$ to $0.5$. 
These axions in the cloud are non-relativistic, oscillating coherently with a frequency approximately equal to the axion mass $m_a$.
A tighter upper bound on $\alpha$ depends on the spin of the black hole, as we will discuss later and in \textbf{Methods}. The lower bound on $\alpha$ is set by requiring the superradiant timescale to be much smaller than the age of the universe \cite{Dolan:2007mj}, i.e., within the range of $10^9$ years in this study. It corresponds to an oscillation period of the axion cloud wavefunction shorter than $20$ days.  
We note that the axions in the mass window studied here are unlikely to be the QCD axion, but rather axion-like particles.}

Due to the axion-photon coupling $g_{a\gamma}aF_{\mu\nu}\tilde{F}^{\mu\nu}/2$ where $F_{\mu\nu}$ is the field strength tensor of the photon and $\tilde{F}^{\mu\nu}$ is the dual tensor, 
the temporal and spatial variations of an axion background field induce a 
change in the dispersion relation of photons. Thus the EVPA of a linearly 
polarized photon, labeled as $\chi$, experiences a rotation due to the 
axion background as
\beq\label{bi-eff} \Delta \chi = g_{a\gamma} [a(t_{\rm obs}, \textbf{x}_{\rm obs}) - a(t_{\rm emit}, \textbf{x}_{\rm emit})],\eeq
depending only on the axion field values at the emission and observation 
points\cite{Carroll:1989vb,Harari:1992ea}. A generalization to curved 
spacetime was discussed in Ref.\,\cite{Schwarz:2020jjh}, reaching the same 
conclusion. We emphasize that this simple expression only holds for photons propagating in vacuum, when the photon wavelength is much smaller than the axion's Compton wavelength. In-medium effects need to be included by solving the radiative transfer equations along the photon path in more realistic cases.

The axion density around the Earth is negligible compared with that of an 
axion cloud surrounding M87$^\star$, and hence can be neglected. Without including in-medium effects, 
for a linearly polarized photon emitted at $(t, r,\theta, \phi)$ in the 
Boyer-Lindquist (BL) coordinates of the black hole, its EVPA shift is 
directly related to the wavefunction of the axion cloud\cite{Chen:2019fsq}
\bea 
\Delta \chi (t, r,\theta, \phi) \approx -  \frac{b c R_{11} (r) \sin \theta \cos{ [\omega t -  \phi]}} {2 \pi R_{11} (r_{\textrm{max}})}.
\eea
For a fixed SMBH, the radial wave-function $R_{11} (r) / R_{11} (r_{\textrm{max}})$ depends on the axion mass. $b\equiv a_{\rm max}/f_a$ is introduced to describe the peak value of the axion cloud; due to nontrivial self-interaction, both numerical simulations\cite{Yoshino:2012kn,Yoshino:2013ofa,Yoshino:2015nsa} and analytic estimates\cite{Baryakhtar:2020gao} indicate that $b\sim O(1)$ in the parameter space we are interested in.  {The parameter $c$ relates the axion-photon 
coupling $g_{a\gamma}$ to the  decay constant $f_a$ via
$c \equiv 2 \pi g_{a \gamma} f_a$, and it is the fundamental quantity that this work constrains.}

In reality, plasma effects also need to be accounted for to properly model the variation of the EVPA. 
In particular, the axion-induced birefringent effect should be combined with astrophysical Faraday rotation. 
As a consequence, the Faraday rotation coefficient obtains an additional contribution, i.e., $\rho_{V} = \rho_{V}^{\textrm{FR}} - 2 g_{a\gamma} \frac{d a}{d s}$ where $s$ is the proper time. The first term $\rho_{V}^{\textrm{FR}}$ describes the frequency-dependent Faraday rotation induced by the plasma effect. The second term is from the axion field, which is an additional frequency-independent contribution along the line-of-sight. The summation of these two effects
characterizes the phase variation in $Q + i U$, where $Q$ and $U$ are the Stokes parameters for the linear polarization.

{In this study, we model the accretion disk by the analytic radiatively inefficient accretion flow (RIAF)\cite{Pu:2018ute} with a sub-Keplerian velocity distribution and vertical  magnetic field \cite{EHTM}. 
The thickness of the accretion flow is parametrized by a dimensionless quantity 
$H \equiv h/R$. Here $h$ is the height scale, and $R$ is the horizontal scale of the accretion flow. We take $H$ as 0.3 and 0.05 for a magnetically arrested disk (MAD)\cite{Igumenshchev:2003rt,Narayan:2003by,McKinney:2012vh,Tchekhovskoy2015}.
The normalization of the electron density is set to be $\sim 10^5$ cm$^{-3}$ so that the total flux density 
at 230 GHz is about $0.5$ 
Jy\cite{Akiyama:2019cqa,Akiyama:2019bqs,Akiyama:2019eap,Akiyama:2019fyp}
and the magnetic field strength is consistent with EHT estimates\cite{EHTM}.}

\begin{figure}[htb]
\centering
\includegraphics[width=0.7\textwidth]{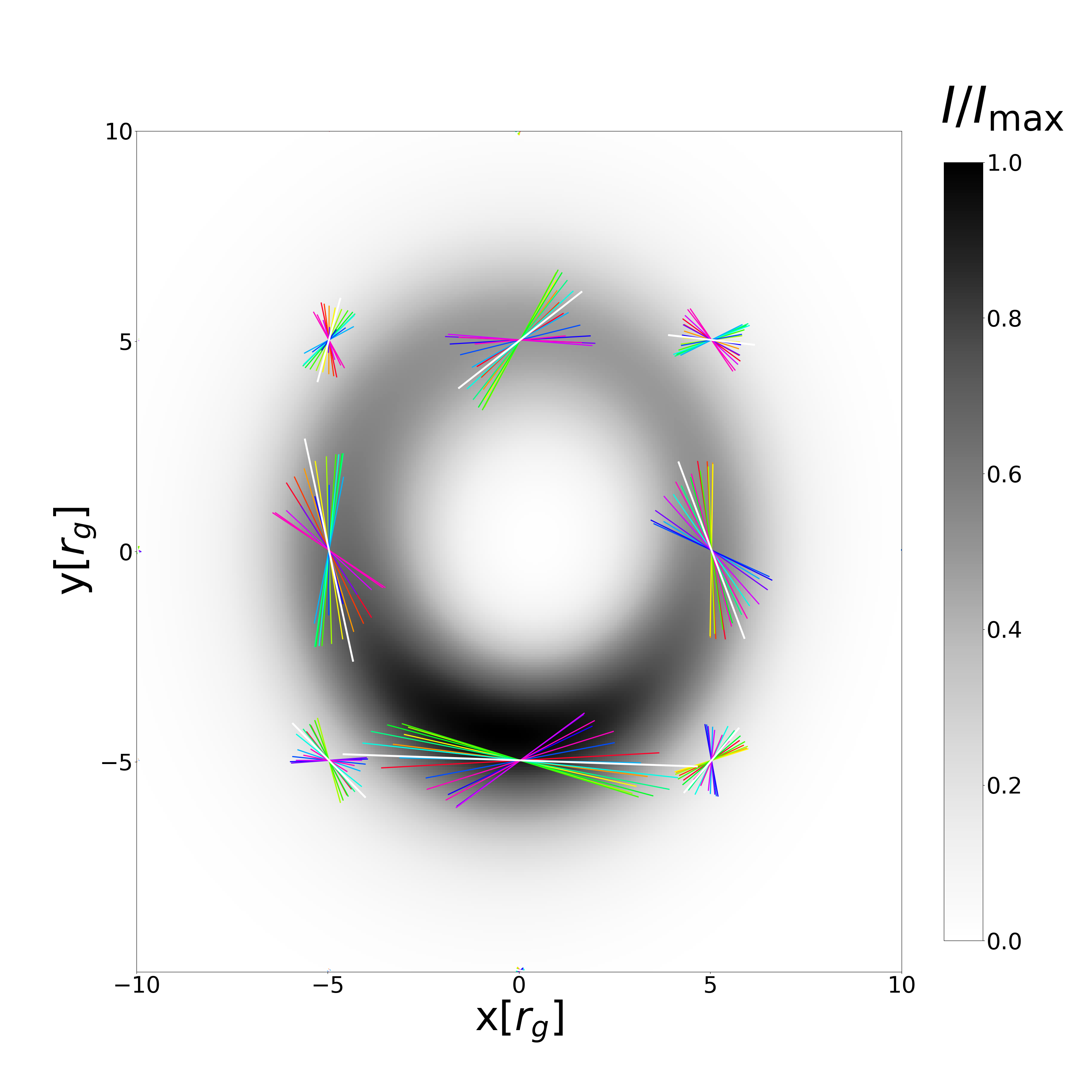}
\caption{{\bf Illustration of the \texttt{IPOLE} simulation of polarized
emission from a Kerr black hole surrounded by an axion cloud.} 
For axion parameters we take $\alpha = 0.4$, and $g_{a\gamma} 
a_{\textrm{max}} = 3\pi/{4}$ as a benchmark. An analytic RIAF with 
$H = 0.3$, vertical magnetic field, and sub-Keplerian velocity 
distribution are assumed. The white quivers are the EVPAs without the 
axion-induced birefringence effect at each point. Different colors, 
ranging from red to purple, represent the EVPA time variation when the 
axion cloud exists. The $+ x$ axis is taken to be aligned with the 
direction of jet projection on this plane.}
\label{exampleAxion}
\end{figure}

{As an illustration, we consider the SMBH M87$^\star$ at a $17^\circ$ inclination angle with respect to the sky plane. The quivers in Fig.\,\ref{exampleAxion} show the EVPA variations using the polarized radiation transfer code \texttt{IPOLE}\cite{Moscibrodzka:2017lcu,Noble:2007zx}.} 
The length of each quiver is 
proportional to the intensity of the linear polarization, i.e., 
$\sqrt{Q^2 + U^2}$, and the direction indicates the polarization direction. 
The white quivers show the values of EVPA without the axion-induced 
birefringence effect. One oscillation period of the axion cloud is equally 
divided into 16 segments, and the color of each quiver, from red to purple, 
represents the time evolution.

\section{Variations of the Azimuthal EVPA Pattern}
{On the sky plane, we use the polar coordinates ($\rho, \varphi$) with the origin at the black hole center.} {The axis with $\varphi = 0$ is taken to be aligned with the direction of jet projection on this plane.}
In order to compare with the data, we follow Ref.\,\cite{EHTP} and calculate the intensity weighted 
average EVPA as a function of $\varphi$ on the sky plane, 
\beq \langle \chi (\varphi) \rangle \equiv \frac{1}{2} \textrm{arg}\Big{(} \langle Q \times I \rangle + i \langle U \times I \ \rangle \Big{)}.\label{defIEVPA}\eeq
The intensity weighted average region is taken to be between 
$\rho_{\textrm{in}} \simeq 3 r_g$ and $\rho_{\textrm{out}} \simeq 8 r_g$, 
according to Ref.\,\cite{Akiyama:2019bqs}.

The axion-induced birefringence effect leads to an oscillation in
$\langle \chi (\varphi) \rangle$. For the $l=m=1$ state, this variation can generically be parametrized as 
\bea \label{ansatz}
\Delta \langle \chi (\varphi) \rangle = - \mathcal{A} (\varphi) \cos{ [\omega t + \varphi + \delta(\varphi)]}.
\eea

\begin{figure}[htb]
\centering
\includegraphics[width=0.7\textwidth]{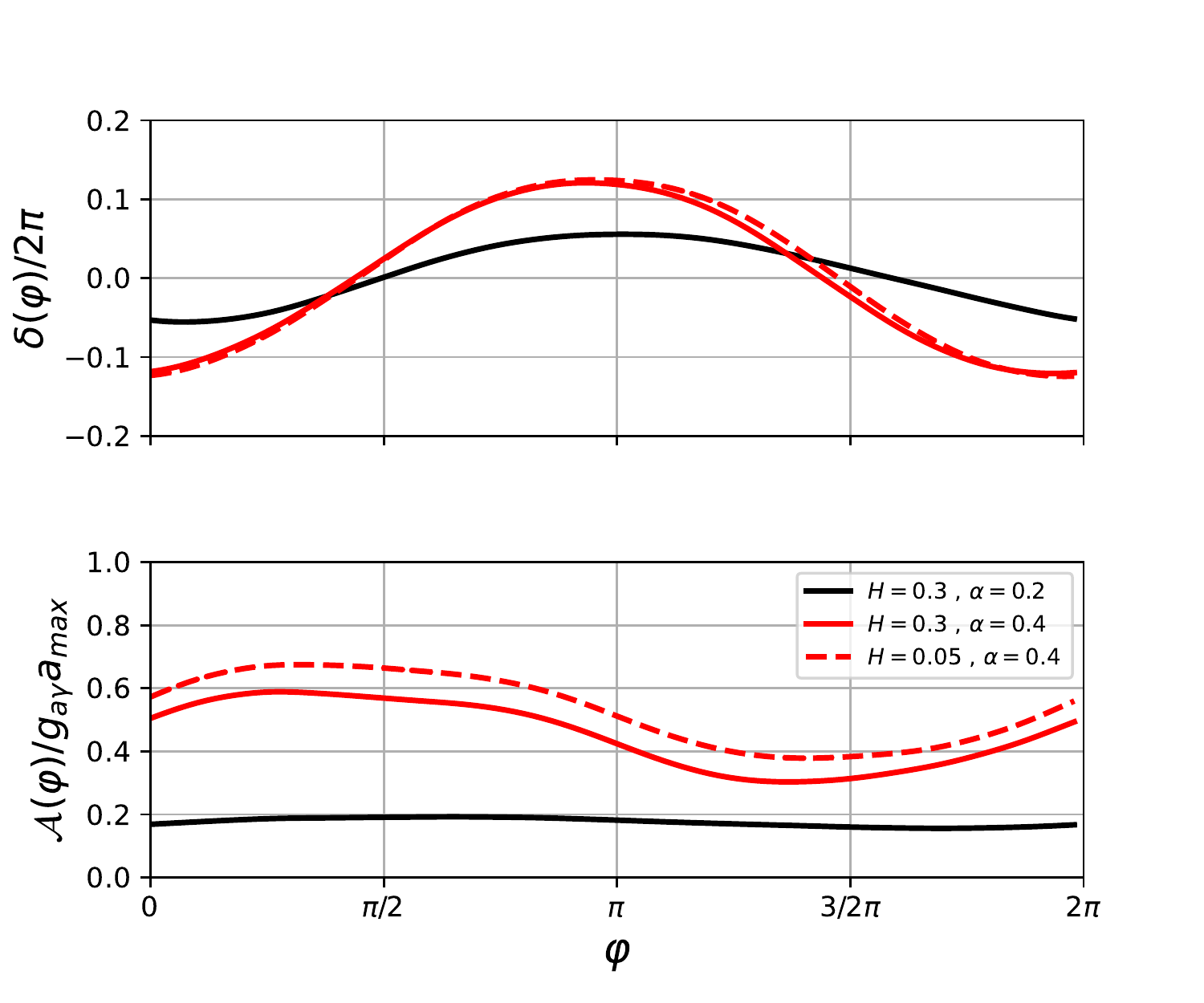}
\caption{{\bf The relative phase $\delta (\varphi)/2\pi$ and amplitude
$\mathcal{A} (\varphi) / g_{a\gamma} a_{\textrm{max}}$ of the 
$\langle \chi (\varphi) \rangle$ variation}. 
The black hole spin is assumed to be $0.99$, and the accretion flow is 
modelled as analytic RIAF using \texttt{IPOLE}, with vertical 
magnetic field and sub-Keplerian velocity distribution. Different 
choices of $H$ and $\alpha$ are shown in the plot. $\varphi = 0$ 
corresponds to the position angle of the jet projection.}
\label{figRIAF0.3}
\end{figure}

In Fig.\,\ref{figRIAF0.3}, we show results from the \texttt{IPOLE} 
calculations with $H = 0.05$ and $H = 0.3$ as two representative models. 
We study the $\varphi-$dependence of the relative phase $\delta$ as well as the 
amplitude $\mathcal{A}$ which is normalized to 
$g_{a\gamma} a_{\textrm{max}} \equiv b c/2\pi$.

We note that $\delta(\varphi)$ is well approximated by a cosinusoidal 
function, specifically
\beq \delta(\varphi) \approx -\omega\ r_{\rm ring}\ \sin{17^\circ}  \cos{\varphi} + \delta_0 \label{phasedelay}.\eeq
where $\delta_0$ is the arbitrary initial phase of the axion cloud.
$r_{\rm ring}$ is approximately the radius of the ring, i.e., $\sim 5\ r_g$, where the dominant emission comes from. This indicates that $\delta(\varphi)$ 
arises from the emission time delay due to the $17^\circ$ 
inclination angle between the M87$^\star$ spin direction and the 
sky plane. In addition, such a fit is also a reliable verification of the mapping between $\varphi$ on the sky plane 
and $\phi$ in BL coordinates, as discussed in \textbf{Methods}.

The behaviour of $\mathcal{A} (\varphi)$ is more subtle, owing to several aspects which must necessarily be considered. 
Firstly, there is an important washout 
effect along the line of sight. The accretion flow is optically thin in 
our study, and thus photons that reach the Earth simultaneously were emitted at different times along the line of sight. Since the axion field oscillates at 
a frequency $\omega \simeq m_a$, these photons experience different axion cloud phases at their time of emission, resulting in a washout effect for $\mathcal{A} (\varphi)$, especially when the radiation size $S_r$ is larger than the Compton wavelength $2\pi\lambda_c$ (see \textbf{Methods} for a more detailed description and Figure\,\ref{washout} for an illustration of a case where emissions and the amplitude of the axion field are constant within a region with size $S_r$). A decrease of $H$ from $0.3$ to $0.05$ reduces this effect slightly, and so we choose $H = 0.3$ as a conservative benchmark for the rest of the discussion.

The other important washout effect comes from the 
lensed photons that experience a much longer propagation time than those 
directly emitted from the accretion flow. The axion-induced EVPA variations for these photons do not add up coherently. For a smaller value of $\alpha$ the Compton wavelength is longer, and the washout effect hence less severe. 
However, $\mathcal{A} (\varphi) / g_{a\gamma} a_{\textrm{max}}$ 
becomes smaller with a smaller $\alpha$ (see the black line in Fig.\,\ref{figRIAF0.3}).  This is mainly because the emission ring is  more spatially separated from the axion cloud peak, and the suppression factor $R_{11} (r) / R_{11} (r_{\textrm{max}})$ becomes smaller. 

Finally, the smearing due to the finite angular resolution of the EHT observation also leads to a washout effect along the $\varphi$-direction\cite{Chen:2019fsq}.

\section{Differential EVPA and Axion Constraints}

The EHT collaboration has released the fiducial polarimetric images from 
four days, i.e., April 5, 6, 10 and 11, 2017\cite{EHTP}. For each of the days, 
there are maps presenting the EVPA distribution as well as the linear 
polarization intensity. In addition, for two of the four days, April 5 
and 11, the intensity weighted EVPA, $\langle \chi (\varphi) \rangle$, 
as functions of the azimuthal angle are also provided. Along the $\varphi$-axis, 
$\langle \chi (\varphi) \rangle$ is obtained by taking the average of a 
wedge with an opening angle of $10^\circ$. 

As discussed previously, we are looking for an overall pattern of 
EVPA variation across the azimuthal angle of the polarized emission. 
It is natural to expect that the astrophysical backgrounds of two sequential days do not change remarkably \cite{EHTP}. Thus we group the 4-day observations into two pairs, (April 5, 6) and (April 10, 11).  For each pair, 
we study the variation of $\langle \chi (\varphi) \rangle$ along the 
azimuthal direction, i.e., $\langle \chi (\varphi, t_j) \rangle - 
\langle \chi (\varphi, t_i) \rangle$ where $t_i$ and $t_j$ represent the 
two sequential days with interval $t_{\textrm{int}} \equiv t_j - t_i = 1$ 
day. To investigate whether there is an evidence of axion-induced birefringent effect we compare the $\langle \chi (\varphi, t_j) \rangle-\langle \chi (\varphi, t_i) \rangle$ obtained in each pair of days with the theoretical prediction due to the axion field.

In the fiducial linear polarization images, there are slight differences 
between the two sequential days. It is not clear whether such differences 
are induced by the intrinsic variation of the accretion flow or by the 
change of the baseline coverage \cite{EHTP}. Thus in our analysis, we 
treat the central value of $\langle \chi (\varphi, t_j) \rangle - \langle 
\chi (\varphi, t_i) \rangle$ as zero. Furthermore, we assume that the 
reconstruction uncertainties for the two sequential days remain the same 
but uncorrelated. Thus the error bar of $\langle \chi (\varphi, t_j) 
\rangle - \langle \chi (\varphi, t_i) \rangle$ is simply that of 
$\langle \chi (\varphi,t_i) \rangle$ multiplied by a factor of $\sqrt 2$.

Meanwhile, $\langle \chi (\varphi, t_j) \rangle - \langle \chi (\varphi, t_i) \rangle$ 
for the two pairs of sequential days are related.
Firstly, ${\mathcal{A} (\varphi)}$ characterizes the difference in the EVPA variation amplitude for two sequential days, and to a good approximation ${\mathcal{A} (\varphi)}$ for the two pairs is the same. 
Furthermore, the phases of the axion field at the times of two pairs 
are related to each other by a relative phase difference of 
($\omega \times$ 5 days). This correlation is accounted for in our analysis.

\begin{figure}[htb]
\centering
\includegraphics[width=0.7\textwidth]{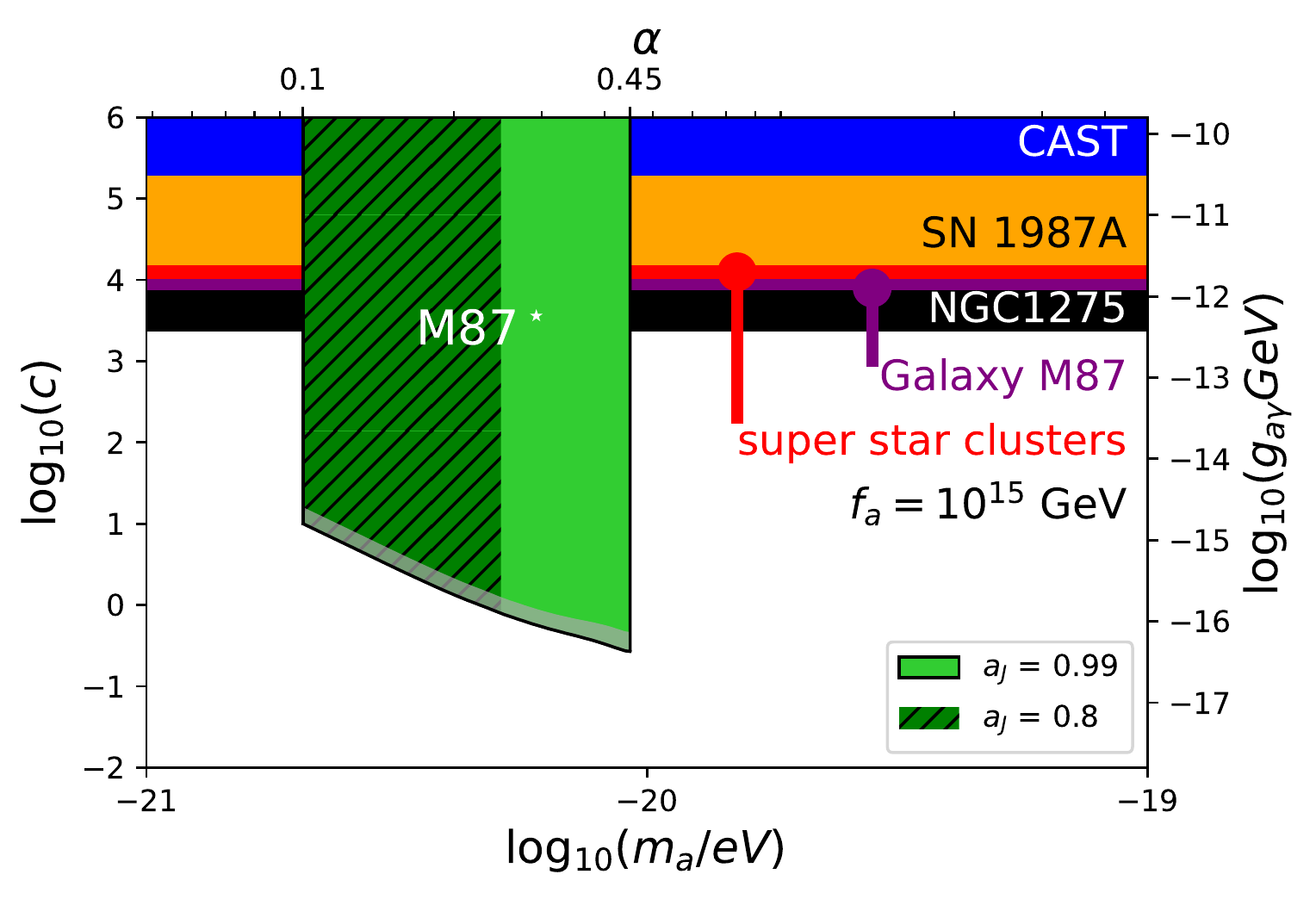}
\caption{{\bf The $95\%$ credible level upper limit (green) on the axion-photon 
coupling, characterized by $c \equiv 2 \pi g_{a \gamma} f_a$, derived 
from the EHT polarimetric observations of SMBH M87$^\star$.}
The black hole spin $a_J$ is assumed to be $0.99$ or $0.80$. The latter case corresponds to a smaller mass window, which overlaps with  $a_J = 0.99$ case in the lower mass region. The gray band at the bottom represents the uncertainty from the five different EVPA reconstruction methods. The bounds from 
CAST\cite{Anastassopoulos:2017ftl}, Supernova 1987A\cite{Payez:2014xsa}, 
super star clusters\cite{Dessert:2020lil}, the M87 Galaxy\cite{Marsh:2017yvc}, 
and NGC1275\cite{Reynolds:2019uqt}, assuming $f_a = 10^{15}$ GeV, 
are shown for comparison.}
\label{figaxionbound}
\end{figure}

For each value of $m_a$, we calculate the likelihood values for 
different $g_{a\gamma} a_{\textrm{max}} \equiv b c/2\pi$. Throughout 
the analysis, the axion field phase $\delta_0$ is treated as a nuisance 
parameter and is marginalized over $[0,2\pi]$. The black hole spin is assumed 
to be either $0.99$ or $0.80$\cite{Tamburini:2019vrf,Feng:2017vba,Davoudiasl:2019nlo}. 
This determines the upper limit of $\alpha$ that we can probe in this study to be $0.44$ or $0.25$ respectively. 
The lower limit of $\alpha$ is taken to be 0.1. With a smaller $\alpha$, the peak of the axion cloud is too far away from the 
photon ring, and the superradiance timescale becomes longer than the age of 
the universe. In Fig.\,~\ref{figaxionbound}, 
 we present the $95\%$ credible level  upper limit on the axion-photon coupling, characterized by the value of $c$. Here we assume that the axion cloud has saturated, 
i.e., $b = 1$. The bound becomes weaker for smaller 
axion masses. This is due to a smaller value of the suppression factor $R_{11} (r) / R_{11} 
(r_{\textrm{max}})$, as well as a smaller 1-day axion field variation due to the longer oscillation period.

{We also compare our constraints with the bounds from 
CAST\cite{Anastassopoulos:2017ftl}, Supernova 1987A\cite{Payez:2014xsa}, 
super star clusters\cite{Dessert:2020lil}, the M87 Galaxy\cite{Marsh:2017yvc}, 
and NGC1275\cite{Reynolds:2019uqt}. While all these existing studies put constraints on $g_{a\gamma}$, our result provides a novel probe to the conversion factor, i.e., $c \equiv 2 \pi g_{a \gamma} f_a$. This is because the axion field saturates the largest possible value in the axion cloud, thereby providing a useful complementarity to the existing landscape of axion searches. 
In terms of the axion models, the value of $c$ is related to the anomaly coefficients in the ultraviolet theory. In addition, an exponentially large value of $c$ can appear in models of the clockwork axion\cite{Kaplan:2015fuy,Farina:2016tgd}. 
In order to directly compare our results with others, we take $f_a$ to be $ 10^{15}$ GeV as a benchmark, motivated by theories of Grand Unification. For the axion mass window 
we consider, a large region of previously unexplored parameter space is covered by the EHT observations. The previous constraints are improved upon
by several orders of magnitude, as shown in Fig.\,\ref{figaxionbound}. We note that our constraints are stronger yet for larger values of $f_a$.  However, they are not necessarily applicable beyond $f_a$ at $10^{16}$ GeV since the axion self-interaction becomes too weak to prevent superradiance\cite{Arvanitaki:2010sy}, and the axion field value does not necessarily saturate.}

\section{Conclusions and Future Prospects}

Polarized imaging of the vicinity of SMBHs offers a unique probe 
to search for axions, thanks to the EVPA oscillation of linearly polarized 
photons arising from the axion-induced birefringence effect. In this study 
we introduced a novel data analysis method that may significantly 
reduce the astrophysical background, and applied it to four 
days of polarization measurements of M87$^\star$ by the EHT\cite{EHTP}. 
EHT observations can as a result rule out a region of the axion mass and axion-photon coupling parameter space, which is unexplored by previous experiments. 

We emphasize that a further optimized search strategy could be applied 
with improved measurements and analysis of the emitted polarization, 
e.g., the upcoming EHT observations with higher time cadence or the Next 
Generation EHT (ngEHT) in the future. Firstly, increasing the overall statistics 
would definitely improve the sensitivity, by incorporating more available 
pairs of sequential days for our differential analysis and 
EVPA reconstruction  with higher precision. 
Further gains can also be made if the results were provided in shorter time segments within one day.  
Furthermore, since the axion-induced birefringence effect is independent of the photon frequency, polarization measurements at different frequency bands such as 345 GHz will be 
extremely helpful to distinguish the axion-induced effect from
Faraday rotation in the plasma. 
The radial distributions of the EVPA are also valuable, and the sensitivity could be improved by using a fuller modelling of the axion profile. 
 Finally, we note that the EVPA observed outside the photon ring is free from the contamination of lensed photons, which give the dominant contribution to the washout effect. Removing these 
lensed photons would help to provide a universal signal prediction, 
independent of accretion flow models. All these 
improvements are within expectations of either the future EHT data release or the ngEHT with an increased baseline coverage\cite{2021ApJS..253....5R}.

\begin{center}
{\bf Methods}
\end{center} 

{\bf Black Hole Superradiance}
Through the superradiance mechanism, a rapidly spinning black hole 
can generate an exponentially growing axion cloud, if the axion has 
a reduced Compton wavelength $\lambda_c$ comparable to the gravitational radius 
of the Kerr black hole $r_g$\cite{Penrose:1971uk,ZS,Press:1972zz,Damour:1976kh,Zouros:1979iw,Detweiler:1980uk,Strafuss:2004qc,Dolan:2007mj} 
(for a review see Ref.\,\cite{Brito:2015oca}). Such a process is terminated 
by either the axion self-interaction or a sufficient loss of the angular 
momentum of the black hole. The wave-function of the axion cloud can be 
written as
\beq a(x^\mu) = e^{-i\omega t} e^{im\phi} S_{lm} (\theta) R_{lm} (r),\eeq
where $x^\mu = [t, r, \theta, \phi]$ are the Boyer-Lindquist (BL) coordinates. 
Such a bound state formed by the black hole and the axion cloud is closely analogous to a hydrogen atom, with the coupling constant being 
$\alpha \equiv r_g/\lambda_c$.  The $\theta$ dependence is characterized 
by the spheroidal harmonics $S_{lm}$ which simplify to $Y_{lm}$ in the 
non-rotating limit of the black hole, or the non-relativistic limit of the 
axion cloud. $R_{lm} (r)$ is the radial component of the wave function 
that vanishes at both the horizon of the black hole and at infinity. 
For a benchmark axion with Compton wavelength satisfying $\alpha = 0.4$, 
the cloud peaks close to the region where the dominant emissions come from\cite{Chen:2019fsq}.
Since the radius at which the axion field reaches the maximum, $r_\textrm{max}$, scales as $1/\alpha^2$, $r_\textrm{max}$ becomes larger for smaller values of 
$\alpha$.

The highest superradiant rate happens for $l = 1, m = 1$, which is the 
lowest energy state amongst those satisfying the superradiance 
condition\cite{Dolan:2007mj}. In this case, for all values of $\theta$, 
the axion cloud wavefunction peaks at the equatorial plane of the black 
hole, i.e., $\sin \theta = 1$. $r_\textrm{max}$ becomes larger for higher modes with a bigger azimuthal number $l$. {To be conservative, we only focus on the $(1,1)$ state and do not consider higher modes in this study.}

The range of the superradiance condition for $\alpha$ is sensitive to the 
black hole spin $a_J$, which is still uncertain for M87$^\star$\cite{EHTP}. 
Refs.\,\cite{Tamburini:2019vrf,Feng:2017vba} claim M87$^\star$ to be a 
nearly extremal Kerr black hole.
From a recent synthetic modeling study of M87 jet at 86\,GHz, higher black hole spin is favored.\cite{Cruz-Osorio:2021}
Thus in this study, we take the black 
hole spin $a_J$ to be 0.99 and 0.8\cite{Davoudiasl:2019nlo} as two 
benchmarks.
{For a fixed azimuthal mode $m$ and black hole spin $a_J$, the superradiance 
condition imposes an upper limit on $\alpha$\cite{Dolan:2007mj},
\beq \alpha<\frac{a_J\ m}{2\ \left(1 + \sqrt{1-a_J^2} \right)},\label{SRC}\eeq
With $m = 1$, $\alpha$ can be at most $0.5$ for an extremal Kerr black hole 
and $0.25$ if $a_J = 0.8$. We set a lower limit on $\alpha $ of $0.1$, so that the superradiant timescale is much shorter than the age of the universe, i.e., within the range of $10^9$ years. The black hole spin can be as low as $a_J = 0.5$ \cite{Dolan:2007mj} in order to satisfy the superradiance condition for $\alpha=0.1$. Once this is met 
the axion cloud profile is only slightly influenced by the value of $a_J$, 
with $\alpha$ being fixed\cite{Amorim:2019hwp}.}

The robustness of the axion cloud is discussed in Ref.\,\cite{Arvanitaki:2009fg},
where several potential effects which may destroy the axion cloud are discussed. 
The tidal forces from a companion star turn out to be negligible. The metric is always dominated by the SMBH, especially for our region of interest. 
One may also be concerned about the possibility of a merger with another SMBH in 
the past, however, we mainly focus on the axion mass range with relatively 
short timescales for superradiance. Even if such a drastic merger happened once 
the axion cloud should have enough time to build up again, and thus we neglect this possibility in our study.

With the growth of the axion cloud, the axion field value gets close to the 
decay constant $f_a$ that governs the self-interaction of axions from the 
axion potential $V(a) = m_a^2 f_a^2 \left( 1 - \cos a/f_a \right)$, where 
$m_a$ is the mass of the axion. Due to nontrivial self-interactions, 
the superradiance is terminated. Consequently the axion cloud enters a 
self-interaction dominated regime, and black hole spin measurements would not apply anymore \cite{Arvanitaki:2010sy,Arvanitaki:2014wva,Brito:2014wla,Davoudiasl:2019nlo,Stott:2020gjj,Unal:2020jiy}.
The fate of the axion cloud can 
either be a violent bosenova or a saturated phase\cite{Yoshino:2012kn,Yoshino:2013ofa,Yoshino:2015nsa,Baryakhtar:2020gao,Omiya:2020vji}. 
Interestingly, in either case, the numerical 
simulations\cite{Yoshino:2012kn,Yoshino:2013ofa,Yoshino:2015nsa} and the 
analytic estimates\cite{Baryakhtar:2020gao} show that the maximum of the 
field value $a_{\rm max}$ remains close to $f_a$ as long as the saturation
regime is reached. This leaves our conclusion insensitive to this subtlety. 
We introduce $b\equiv a_{\rm max}/f_a$ to describe the peak value of the 
axion cloud. We note that in Ref.\,\cite{Baryakhtar:2020gao}, the value of $b$ is shown 
to decrease at lower $\alpha$ and logarithmically dependent on the mass of 
the black hole.

{\bf Accretion Flow and Radiative Transfer}
For low-luminosity active galactic nuclei, such as Sgr A$^\star$ and 
M87$^\star$, the accretion flow are approximately described as radiatively 
inefficient accretion flow (RIAF)\cite{Narayan:1996wu,Yuan:2014gma}, 
which are geometrically thick and optically thin at 230 
GHz\cite{Yuan:2014gma,Prieto:2015efa, Akiyama:2019fyp}. The accretion
flow thickness is characterized by a dimensionless quantity 
$H \equiv h/R$ where $h$ is the height scale and $R$ is the horizontal scale of the accretion flow \cite{Pu:2018ute}. 
For a magnetically arrested disk (MAD), which 
well-describes M87$^\star$\cite{EHTM}, $H$ is compressed to be $0.05$ by 
the strong magnetic field in the inner region ($\lesssim 10\ r_g$) and 
becomes $\sim 0.3$ in the outer region\cite{Igumenshchev:2003rt,Narayan:2003by,McKinney:2012vh,Tchekhovskoy2015}.  
In this study we adopt the analytic RIAF model \cite{Pu:2018ute} as a 
benchmark model, and vary $H$ in order to understand the uncertainties 
induced by the thickness of the accretion flow. As demonstrated later, 
a different choice of $H$ does not change our conclusion qualitatively.

The Stokes parameters \((I, Q, U, V)\) are generically applied to describe 
the properties of macroscopic polarization (see e.g., 
Ref.\,\cite{2014JKAS...47...15T}), in which $I$ is the total intensity, 
$Q$ and $U$ characterize the linear polarization, and $V$ describes the 
circular polarization. The in-medium effects lead to mixtures among the four 
Stokes components in the polarized radiative transfer equation
\beq
\frac{d}{d s}\left(\begin{array}{l}I \\ Q \\ U \\ V\end{array}\right)=\left(\begin{array}{l}j_{I} \\ j_{Q} \\ j_{U} \\ j_{V}\end{array}\right)-\left(\begin{array}{llll}\alpha_{I} & \alpha_{Q} & \alpha_{U} & \alpha_{V} \\ \alpha_{Q} & \alpha_{I} & \rho_{V} & \rho_{U} \\ \alpha_{U} & -\rho_{V} & \alpha_{I} & \rho_{Q} \\ \alpha_{V} & -\rho_{U} & -\rho_{Q} & \alpha_{I}\end{array}\right)\left(\begin{array}{l}I \\ Q \\ U \\ V\end{array}\right),\eeq
where $s$ is the proper time, \(j_{I, Q, U, V}\) are the polarized 
emissivities, \(\alpha_{I, Q, U, V}\) are the absorption coefficients, 
and \(\rho_{Q, U, V}\) are the Faraday rotation and conversion coefficients.

The axion-induced birefringence effect can be properly included in the 
radiative transfer matrix as  
\beq\label{ipole-mod} \rho_{V} = \rho_{V}^{\textrm{FR}} - 2 g_{a\gamma} \frac{d a}{d s},\eeq 
where the first term is the coefficient of the frequency-dependent Faraday 
rotation from the plasma, and the second term is from the axion field which 
gives a frequency-independent addition along the line-of-sight. $\rho_{V}$ 
characterizes the rotation of the phase in $Q + i U$\cite{Carroll:1989vb,Harari:1992ea,Plascencia:2017kca,Ivanov:2018byi,Fujita:2018zaj,Liu:2019brz,Fedderke:2019ajk,Caputo:2019tms,Yuan:2020xui}, which is related to 
the EVPA through
\beq \chi \equiv \frac{1}{2} \mathrm{arg} (Q + i U).\eeq 

To properly model the axion-induced birefringence effect, we modify the 
polarized radiative transfer equation in the general relativistic radiation 
transfer code \texttt{IPOLE}\cite{Moscibrodzka:2017lcu,Noble:2007zx}, 
as indicated in Eq.\,(\ref{ipole-mod}), with analytic RIAF models in which 
$H = 0.05$ and $0.3$ are implemented. For the velocity distribution, we choose 
the sub-Keplerian flow as an approximation. The magnetic field is taken 
to be vertical as a benchmark model\cite{EHTM}.  In addition, we adjust 
the normalization of the electron density to be $\sim 10^5$ cm$^{-3}$ 
compared to the original model so that the total flux density in the image 
at 230 GHz is about $0.5$ 
Jy\cite{Akiyama:2019cqa,Akiyama:2019bqs,Akiyama:2019eap,Akiyama:2019fyp}
and the magnetic field strength is consistent with the EHT estimate\cite{EHTM}. We also perform a more comprehensive study in order to investigate the effects of various choices in the RIAF model, such as the direction of the magnetic field (e.g., toroidal or radial) and the velocity distributions. We find that varying these parameters does not affect our conclusion qualitatively.

The accretion flow, on the other hand, may also be affected by the black hole 
spin.  Taking sub-Keplerian RIAF as the benchmark model, we compare the 
axion cloud induced EVPA time variations with different choices of the 
black hole spin, $a_J = 0.5$ and $a_J = 0.8$. We set the value of $\alpha$ to be close to 0.1 in order 
to  satisfy the superradiance condition Eq.\,(\ref{SRC}). We find that the EVPA variations remain qualitatively the same as those in the scenario with $a_J = 0.99$.
This is due to the fact that the line-of-sight washout effect becomes 
negligible for a lower value of $\alpha$, i.e., longer Compton wavelength. More details will be discussed later in the \textbf{Line-of-Sight Washout Effect} Section in \textbf{Methods}. 

{In order to compare with EHT data, we need to map the sky plane coordinates ($\rho, \varphi$)} to the BL coordinates of the SMBH.
This can be done through ray-tracing in \texttt{IPOLE}. For an MAD, 
the dominant emission comes from the region near the equatorial plane 
of the black hole. Combined with the fact that the brightest region is 
around $5 \ r_g$ away from the center of the black hole, for SMBH M87$^\star$ whose orientation is almost face-on, $\varphi$ can be directly mapped to $-\phi$ 
in BL coordinates to a good approximation. The minus sign is due 
to the fact that the spin direction of M87$^\star$ points away from the 
Earth, i.e., M87$^\star$ rotates clockwise on the sky plane. In addition, 
for the region of interest in this study $\rho$ can be approximately 
mapped to $r$ in BL coordinates as well; $\rho \simeq r$. 

There are 
also photons that propagate around the black hole several times before 
reaching the Earth due to lensing effects\cite{Johannsen:2010ru, 
Gralla:2019xty, Johnson:2019ljv, Gralla:2019drh}, which can influence the 
polarization as well\cite{Jimenez-Rosales:2021ytz}. For the total 
intensity they contribute subdominantly, providing $\sim 10$\% at most. 
However, if one uses the intensity weighted EVPA 
as the observable, their contribution is non-negligible and can lead to 
a noticeable washout of the axion-induced birefringence, which contributes to the  asymmetry along $\varphi$ in Fig.\,\ref{figRIAF0.3}. We note that 
such a washout effect can be largely removed if the detailed EVPA 
distribution is provided at different $\rho$.

{\bf Data Characterization}
To calculate $\langle \chi (\varphi, t_j) \rangle - \langle \chi 
(\varphi, t_i) \rangle$, one has to take into account the 
observation time in each day, i.e., $t_{\textrm{obs}}\simeq 6$ hours. 
Using the parametrization in Eq.\,(\ref{ansatz}), we have 
\bea 
&&\langle \chi (\varphi, t_j) \rangle - \langle \chi (\varphi, t_i) \rangle \nn\\ &=& \frac{\mathcal{A}(\varphi) }{t_{\textrm{obs}}} \int^{t_{\textrm{obs}}/2}_{-t_{\textrm{obs}}/2} \{ \cos{ [\omega (t_i + \delta t) + \varphi + \delta(\varphi)]} \nn\\ &-& \cos{ [\omega (t_j + \delta t) + \varphi + \delta(\varphi)]} \} \ d \delta t\nn\\
&=& \widetilde{\mathcal{A} (\varphi)} \sin{[\omega (t_i + t_{\textrm{int}}/2) +  \varphi + \delta(\varphi)]}\label{deltaEVPA}\eea
where $\widetilde{\mathcal{A} (\varphi)} = 2 \mathcal{A} (\varphi) \sin{[\omega t_{\textrm{int}}/2]} \sin{[\omega t_{\textrm{obs}}/2]} / (\omega t_{\textrm{obs}}/2)$. 
The factor $\sin{[\omega t_{\textrm{obs}}/2]} / (\omega t_{\textrm{obs}}/2)$ 
represents the washout effect due to the time averaging over the observation 
time of each day, while $\sin{[\omega t_{\textrm{int}}/2]}$ represents 
how much the axion cloud has changed during the time interval of a given day. 
For the parameter space we are interested in, the axion oscillation period 
is larger than one day. Thus $\sin{[\omega t_{\textrm{int}}/2]}$ contributes 
as a suppression factor, which makes the constraints in the lower mass region weaker in Fig.\,\ref{figaxionbound}. Finally, we note that one can absorb
$\omega (t_i + t_{\textrm{int}}/2)$ into the phase of the axion cloud 
$\delta_0$, which is a nuisance parameter in the analysis.

In Fig.\,8 of Ref.\,\cite{EHTP}, $\langle \chi (\varphi) \rangle$ is
constructed using five different methods. The obtained results do not 
perfectly agree with each other, reflecting possible differences in the systematic uncertainties of each analysis. However, we are mainly 
interested in the time variation of the EVPA. We therefore assume that these 
systematics can be largely eliminated for this differential study.

The reconstruction of $\langle \chi (\varphi) \rangle$ from the data 
is subtle. For example, there are nontrivial leakages between the 
polarization modes in the measurements, characterized by the so-called D-term. 
These can lead to uncertainties in $\langle \chi (\varphi) \rangle$. 
For some values of $\varphi$ (particularly when the polarized intensity 
is low), the reconstruction gives ambiguous results, which appear as 
bifurcations in the figure. 
In our analysis we apply a conservative approach, removing the range of $\varphi$ where reconstruction ambiguities appear. 
We choose a bin-size of $10^\circ$ in order to reduce the 
correlations among different azimuthal angles. 
We only keep the bins 
whose EVPA distribution can be properly fit by a Gaussian function without ambiguous bifurcations.
For example, in the results produced by the \texttt{polsolve} method\cite{Mart_Vidal_2021}, the left bins have azimuthal angles ranging between 30$^\circ$ to 90$^\circ$ and 170$^\circ$ to 330$^\circ$ for the April 5 
data, and 30$^\circ$ to 310$^\circ$ for the April 11 data. Effectively 
we then have $53$ bins in total. For each bin, we fit the stripe width by a 
Gaussian function to obtain an estimation of the measurement error. 
The errors roughly range from $\pm 3^\circ$ to $\pm 15^\circ$. For other methods the uncertainties in the regions with strong intensity are nearly the same, and thus they give comparable axion constraints as those shown in Fig.\,\ref{figaxionbound}.  

{\bf Statistics} {
We use Bayesian statistics to derive the 95\% credible level (C.L.) upper limits on the axion-photon coupling, characterized by $c$, for different axion masses $m_a$ (or equivalently, $\alpha$, for a given black hole mass). The marginalized posterior of $c$ is written as
\begin{equation}
    \Pr\left(\log_{10}c|\boldsymbol{d},m_a\right) = \frac{1}{\rm Const}\int_0^{2\pi} \frac{{\rm d}\delta_0}{2\pi}\prod_i\frac{1}{\sqrt{2\pi}\sigma_i}e^{-\left(d_i-d^0_i(c,m_a,\delta_0)\right)^2/\left(2\sigma_i^2\right)},\label{posterior}
\end{equation}
where we assume that the priors of the initial phase $\delta_0$ and $\log_{10}c$ are uniform. The denominator, ${\rm Const}$, is the normalization factor which does not need to be calculated in our case. In the exponent $d_i$ represents the EVPA variation in a sequential-day pair in the experimental data, while the sub-index $i$ runs over both the azimuthal angle bins and the two pairs, i.e., $\langle \chi (\varphi,{\rm day}6) \rangle - \langle \chi (\varphi,{\rm day5}) \rangle $ and $\langle \chi (\varphi,{\rm day11}) \rangle - \langle \chi (\varphi,{\rm day10}) \rangle $. The parameter $d_i^0(c,m_a,\delta_0)$ is the theoretical prediction of the EVPA variation, derived from Eq.\,(\ref{deltaEVPA}), while $\sigma_i$ characterizes the uncertainty in the data. As stated in Ref.\,\cite{EHTP}, no significant EVPA variations are observed in any sequential-day pairs, which indicates ${d_i}=0$. We then perform the integration on the marginalized posterior in order to obtain the cumulative probability of $\log_{10}c_{\rm ul}$, where $c_{\rm ul}$ is the upper limit of c to achieve a corresponding probability value,
\begin{equation}
    \Pr\left(\log_{10}c<\log_{10}c_{\rm ul}|\boldsymbol{d}=0,m_a\right) =  \int_{\log_{10}c_{\rm min}}^{\log_{10}c_{\rm ul}}\Pr\left(\log_{10}c|\boldsymbol{d}=0,m_a\right){\rm d}\log_{10}c.
\end{equation}
Here we choose $c_{\rm min}=\alpha_{\rm EM}\simeq1/137$, given by the scenario with one species of charged fermion contributing to this anomaly vertex. Here $\alpha_{\rm EM}$ is the electromagnetic fine structure constant. A lower value of $c_{\rm min}$ is possible and it will lead to tighter constraints. The 95\% C.L. upper limit for $\log_{10}c$, denoted as $\log_{10}c_{\rm 95\%}$, is calculated by inversely solving the following equation
 \begin{equation}
     95\% = \frac{\Pr\left(\log_{10}c<\log_{10}c_{\rm 95\%}|\boldsymbol{d}=0,m_a\right)}{\Pr\left(\log_{10}c<\log_{10}c_{\rm max}|\boldsymbol{d}=0,m_a\right)}
 \end{equation}
where we choose $c_{\rm max}$ to be a large number, such as $10^{3}$. We note that our result is not sensitive to this choice, and the normalization factor ${\rm Const}$ is also automatically cancelled in this calculation.

Of course, there are uncertainties from the time-dependent astrophysical 
background. The time variation of the accretion flow is not well 
understood, and consequently it is not included in our simulation 
discussed above. This leads to the major subtlety when we compare 
results from simulations with observations. For M87$^\star$ 
with mass $6.5\times 10^9$ solar 
masses\cite{Akiyama:2019cqa,Akiyama:2019bqs,Akiyama:2019eap,Akiyama:2019fyp}, 
the typical time scale is 5 days for light propagating near the innermost 
stable circular orbit (ISCO). In addition, for gas in the accretion flow 
with a sub-Keplerian velocity distribution, the associated time scale is 
approximately one month at $r \simeq 5 r_g$. Thus it is reasonable to 
expect that the astrophysical features at large scales, comparable to 
the size of the accretion flow, remain approximately unchanged over a
time scale of one day. The differential EVPA for two sequential days, i.e., 
$\langle \chi (\varphi, t_j) \rangle - \langle \chi (\varphi, t_i) \rangle$, 
is an ideal observable with the astrophysical uncertainties being highly 
suppressed.

To qualitatively estimate this uncertainty among the azimuthal angle 
bins selected in our analysis, we find the common ones shared in April 5 
and April 11 data, and calculate the differences between the average 
values of $\langle \chi (\varphi) \rangle$. Assuming the accretion flow 
dynamics lead to an approximately linear variation during the 6-day 
interval, we divide the difference in each bin by a factor of 6.  This 
provides a rough estimate of the variation induced by the astrophysical 
background. By comparing it with the error bar derived for 
$\langle \chi (\varphi, t_j) \rangle - \langle \chi (\varphi, t_i) \rangle$, 
we find that the astrophysical background uncertainty is generally smaller 
than that from the EVPA reconstruction. This justifies the validity of 
our proposed analysis strategy.

{\bf Line of Sight Washout Effect} 
Here we investigate in detail how the integral along the line of
sight leads to a washout effect. As discussed previously, the axion-induced birefringence can be incorporated via a modification of the radiative transfer matrix. To gain some intuition, 
we simplify the problem by ignoring in-medium effects, keeping only
the source terms and the axion effect. Then the evolution of the linear 
polarization can be written as
\beq \frac{d\Big(Q + i\ U\Big)}{ds} = j_Q + i\ j_U - i 2 g_{a\gamma} \frac{d a}{d s} \Big(Q + i\ U\Big).\eeq
This can be generally solved via
\bea\label{washout1} &&Q (s_f) + i\ U (s_f) =\nn\\ &&\int_{s_i}^{s_f} e^{i 2 g_{a\gamma} \Big(a(s_f) - a(s)\Big)} \Big( j_Q (s) + i\ j_U (s) \Big) ds,\eea
where $s_i$ and $s_f$ are the initial and final points along the line 
of sight. Further simplifying, we assume the radiation source terms, 
$j_Q$ and $j_U$, are constant in a finite region. Thus the washout effect 
is characterized by the axion-dependent integral in Eq.\,(\ref{washout1}). 
For a qualitative estimation, we take $a(s)$ to be a coherently oscillating 
background with a constant amplitude $a_{0}$ over the same region. 
The resulting washout effect on $\mathcal{A}$, defined in Eq.\,(\ref{ansatz}), is shown in 
Figure\,\ref{washout}, as a function of the size of the radiation 
source $S_r$, normalized by the axion Compton wavelength $2\pi \lambda_c$.  
Here we see that the washout effect is negligible if $\lambda_c\gg S_r$, 
and it becomes sizable when these two scales are comparable.

\begin{figure}
\centering
\includegraphics[width=0.6\textwidth]{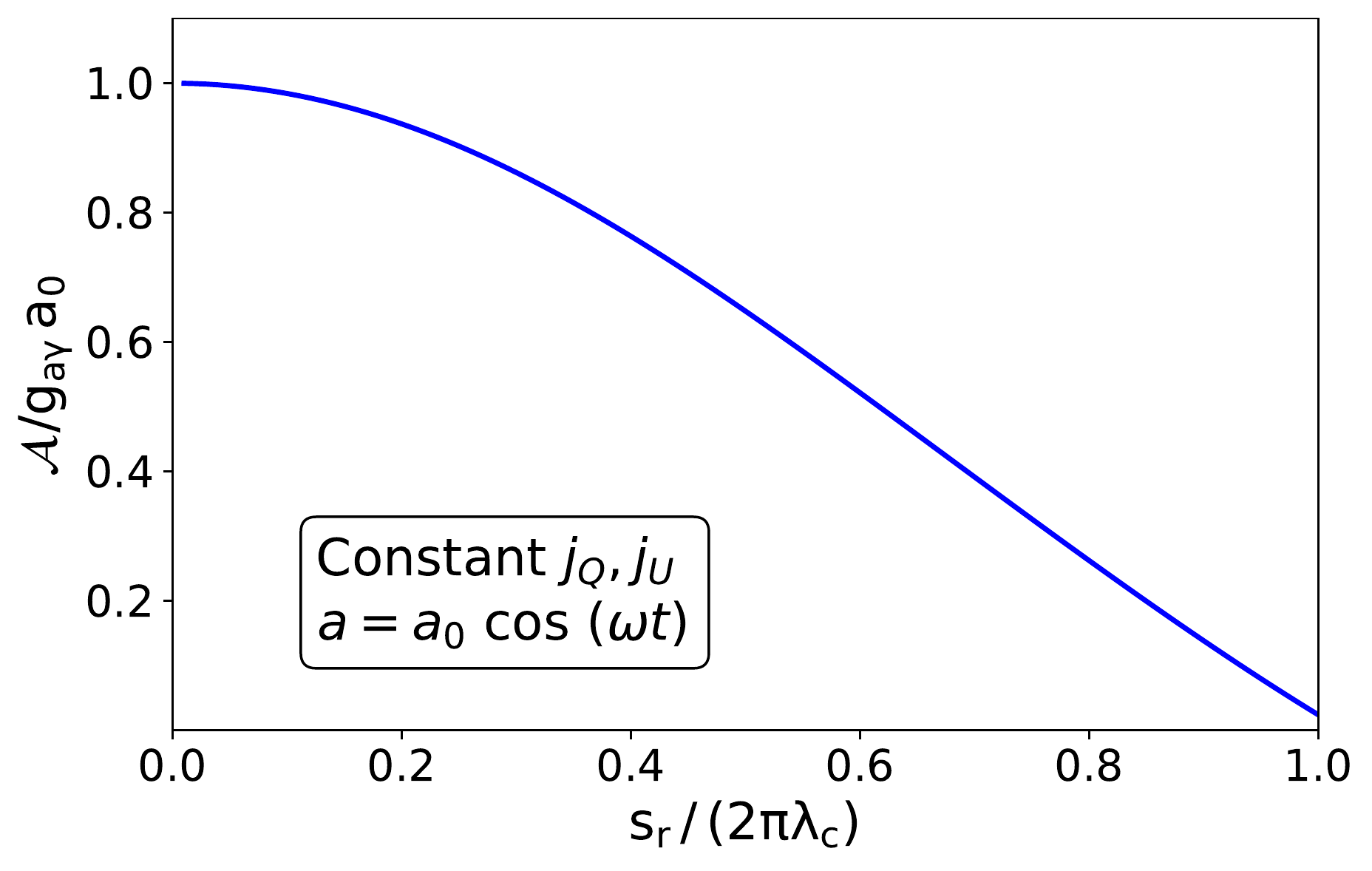}
\caption{{\bf Washout effect on the oscillation amplitude of the EVPA as a function 
of the size of the radiation source $S_r$, normalized by the axion Compton wavelength $2\pi \lambda_c$.} We assume the radiation source terms, $j_Q$ and $j_U$, are constant in a finite region and take the axion field to be a coherently oscillating background with a constant amplitude $a_{0}$ over the region. 
}
\label{washout}
\end{figure}

In a realistic scenario like the optically thin accretion flow around 
M87$^\star$, $S_r$ is determined by the geometric thickness of the 
accretion flow. The washout effect is not important if the accretion 
flow satisfies $r_{\textrm{ring}} H < 2\pi \lambda_c$. Taking 
$\alpha \equiv r_g/\lambda_c = 0.4$ and $r_{\textrm{ring}} \simeq 5 r_g$, 
this condition becomes $H \ll 3$, which is easily satisfied in our RIAF 
model. For smaller $\alpha$, due to the much larger $\lambda_c$, this 
washout effect along the line of sight is further reduced.

Furthermore, there are also contributions from lensed photons; since the dominant emission comes from around the equatorial plane, the observed photon ring contains radiation which has propagated around the black hole several times. 
Since the axion cloud had different phases when these photons were emitted, they also lead to washout effects to the EVPA variations, as shown in Fig.\,\ref{figRIAF0.3}. 
 We note that in more realistic cases, such as the accretion flow in the general relativistic magnetohydrodynamic simulations, the lensed photons are typically less polarized than the direct emission due to magnetic turbulence \cite{Jimenez-Rosales:2021ytz}. Thus using the analytic RIAF as a benchmark in this study serves as a conservative estimation.


\begin{addendum}

\item[Acknowledgements]
We are grateful to Nick Houston, Samuel Liebersbach and Dimitrios Psaltis for careful reading and useful comments on the manuscript, and Chunlong Li, Ye-Fei Yuan, Shan-Shan Zhao and Zihan Zhou for useful discussions. 
Y.C. is supported by the China Postdoctoral Science Foundation under Grants 
No. 2020T130661, No. 2020M680688, the International Postdoctoral Exchange 
Fellowship Program, and by the National Natural Science Foundation of China 
(NSFC) under Grants No. 12047557. 
R.-S.L. is supported by the Max Planck Partner Group of the MPG and the 
Chinese Academy of Sciences (CAS), the NSFC under Grant No. 11933007, the Research Program of Fundamental and Frontier Sciences of CAS 
under Grant No. ZDBS-LY-SLH011, and the Shanghai Pilot Program for Basic Research ? Chinese Academy of Science, Shanghai Branch (JCYJ-SHFY-2021-013).
Y.M. is supported by the ERC Synergy Grant ``BlackHoleCam: Imaging the 
Event Horizon of Black Holes'' under Grant No. 610058. 
J.S. is supported by the NSFC under Grants No. 12025507, No. 11690022, 
No. 11947302, by the Strategic Priority Research Program and Key Research 
Program of Frontier Science of CAS under Grants No. XDB21010200, 
No. XDB23010000, and No. ZDBS-LY-7003 and CAS project for Young Scientists in Basic Research YSBR-006.
Q.Y. is supported by the NSFC under Grants No. 11722328, No. 11851305, by the Key Research Program 
of CAS under Grant No. XDPB15, and by the Program for Innovative Talents and Entrepreneur in Jiangsu. 
Y.Z. is supported by U.S. Department of Energy under Award No. DESC0009959.
Y.C. would like to thank the SHAO and TDLI for their kind hospitality.
Y.Z. would like to thank the ITP-CAS for their kind hospitality.

\item[Author Contributions] JS, QY and YZ initiated this study, YC, XX and YZ developed the method, XX, YL and YC analyzed the data with important contribution from YM and RSL, YM offered  guidance on accretion flow models, YC and YZ wrote the initial draft, with contributions from QY and JS. All authors have reviewed, discussed, and commented on the modeling, data analysis, and the manuscript. Correspondence and requests for materials should be addressed to
YC (yifan.chen@itp.ac.cn), YM (mizuno@sjtu.edu.cn), JS (jshu@itp.ac.cn), QY (yuanq@pmo.ac.cn), and YZ (zhaoyue@physics.utah.edu).

\item[Data Availability] 
The data of polarimetric measurements used in this paper is drawn from the
publicly available publication of Event Horizon Telescope collaboration \cite{EHTP}.
The data that support the plots within this paper and other findings of this study can be found at https://github.com/XueXiao-Physics/Axion\_EHT\_2021.

\item[Code Availability] The simulation codes used in this study are a modified version of publicly available code \texttt{IPOLE} \cite{Moscibrodzka:2017lcu,Noble:2007zx} (https://github.com/moscibrodzka/ipole).
The data analysis codes can be found at\\ https://github.com/XueXiao-Physics/Axion\_EHT\_2021.

\item[Competing Interests Statement] The authors declare no competing interests.
\end{addendum}

\renewcommand{\refname}{References}

\newpage

\renewcommand{\figurename}{Extended Figure}
\renewcommand{\tablename}{Extended Table}
\setcounter{figure}{0}
\setcounter{table}{0}

\end{document}